\RequirePackage[mathlines]{lineno}
\documentclass[prd,twocolumn,showpacs,aps,prl]{revtex4-2}
\usepackage{overpic,graphicx}
\usepackage{dcolumn}
\usepackage{bm}
\usepackage{rotating}
\usepackage{subfigure}
\usepackage{color}
\usepackage[bookmarksnumbered, pdfstartview=FitH,colorlinks,urlcolor=blue, citecolor=blue,linkcolor=blue,] {hyperref}
\usepackage{lineno}
\usepackage{multirow}
\usepackage{amsmath}
\usepackage{makecell}

\begin{document}

\title{\bf \boldmath
Study of $e^+e^-\rightarrow\Omega^{-}\bar\Omega^{+}$ at center-of-mass
energies from 3.49 to 3.67 GeV
}
\author{
\begin{small}
\begin{center}
M.~Ablikim$^{1}$, M.~N.~Achasov$^{11,b}$, P.~Adlarson$^{70}$, M.~Albrecht$^{4}$, R.~Aliberti$^{31}$, A.~Amoroso$^{69A,69C}$, M.~R.~An$^{35}$, Q.~An$^{66,53}$, X.~H.~Bai$^{61}$, Y.~Bai$^{52}$
% ...
\\
\vspace{0.2cm}
(BESIII Collaboration)\\
\vspace{0.2cm} {\it
$^{1}$ Institute of High Energy Physics, Beijing 100049, People's Republic of China\\
% ...
$^{4}$ Bochum Ruhr-University, D-44780 Bochum, Germany\\
% ...
$^{11}$ G.I. Budker Institute of Nuclear Physics SB RAS (BINP), Novosibirsk 630090, Russia\\
% ...
$^{31}$ Johannes Gutenberg University of Mainz, Johann-Joachim-Becher-Weg 45, D-55099 Mainz, Germany\\
% ...
$^{35}$ Liaoning Normal University, Dalian 116029, People's Republic of China\\
% ...
$^{52}$ Southeast University, Nanjing 211100, People's Republic of China\\
$^{53}$ State Key Laboratory of Particle Detection and Electronics, Beijing 100049, Hefei 230026, People's Republic of China\\
% ...
$^{61}$ University of Jinan, Jinan 250022, People's Republic of China\\
% ...
$^{66}$ University of Science and Technology of China, Hefei 230026, People's Republic of China\\
% ...
$^{69A}$ University of Turin and INFN, I-10125, Turin, Italy\\
$^{69B}$ University of Eastern Piedmont, I-15121, Alessandria, Italy
$^{69C}$; (C)INFN, I-10125, Turin, Italy\\
$^{70}$ Uppsala University, Box 516, SE-75120 Uppsala, Sweden\\
% ...
$^{b}$ Also at Novosibirsk State University, Novosibirsk, 630090, Russia\\
% ...
}\end{center}

\vspace{0.4cm}
\end{small}
}
\begin{abstract}
Using data samples of $e^+e^-$ collisions collected with the BESIII
detector at eight center-of-mass energy points between 3.49 and 3.67
GeV, corresponding to an integrated luminosity of 670 pb$^{-1}$, we
present the upper limits of Born cross sections and the effective form
factor for the process $e^+e^-\rightarrow\Omega^{-}\bar\Omega^{+}$. A
fit to the cross sections using a perturbative QCD-derived energy-dependent
function shows no significant threshold effect. The upper limit on the
measured effective form factor is consistent with a theoretical
prediction within the uncertainty of 1$\sigma$. These results provide
new experimental information on the production mechanism of $\Omega$.
\end{abstract}


\maketitle

\oddsidemargin  -0.2cm
\evensidemargin -0.2cm


\section{Introduction}
The $\Omega^-$ resonance, composed of three valence strange quarks, is one of the most famous hyperons. Its discovery was guided directly by the eightfold way~\cite{eightfold_model} more than half a century ago. Until now, its physical properties and inner structure are still not well understood. One approach to parametrize its inner structure is via electromagnetic form factors which can be accessed experimentally with electron-positron annihilation into a virtual photon. For a spin-$S$ baryon, there are $2S+1$ form factors to describe the $\gamma^{*}$-baryon-antibaryon vertex in the electron-positron annihilation under the assumption of one photon exchange. For baryons with $S = \frac{1}{2}$, the two form factors are the magnetic form factor $|G_{M}(q^{2})|$ and the electric form factor $|G_{E}(q^{2})|$. For baryons with $S = \frac{3}{2}$, like the $\Omega^-$ hyperon, there are four form factors to describe the $\gamma^{*}\Omega^{-}\bar{\Omega}^{+}$ vertex, corresponding to electric charge ($|G_{E0}|$), magnetic dipole ($|G_{M1}|$), electric quadrupole ($|G_{E2}|$) and magnetic octupole ($|G_{M3}|$), respectively~\cite{koerner}. The individual form factors can be determined through the analysis of the angular distribution of the hyperon in the electron-positron c.m. frame. Due to limited statistics, most experiments have measured only a combination of the electromagnetic form factors, the so-called effective form factor $|G_{\text{eff}}|$, which can be extracted from the total cross section of the pair production in electron-positron annihilation~\cite{denig, pacetti,  BESIII:2019cuv,  BESIII:2021ccp,Wang:2022zyc, Wang:2022bzl}. A recent calculation of the effective form factor has been performed for the $\Omega^{-}$ hyperon using the covariant spectator quark model~\cite{ramalho}, in which the data from the CLEO-c experiment~\cite{cleo} in the timelike region was used to fix the free parameters.

In addition, many experimental studies determining the cross section of baryon-antibaryon pairs observed an unusual behavior near the threshold. The cross sections for the production of neutron~\cite{nnbar, FENICE, SND2012}, proton~\cite{pbabar, pcmd3, ppbar1, besTag}, $\Lambda$~\cite{BESIII:2017hyw}, and $\Lambda_{c}$ pairs~\cite{LcLcbar} are approximately constant. The only data for the process $e^+e^-\rightarrow\Omega^{-}\bar\Omega^{+}$ were gathered by the CLEO-c experiment at $q^{2}\geq14.2~\text{GeV}^{2}$~\cite{cleo}, which is far from the threshold of $\Omega^{-}\bar\Omega^{+}$ at $q^2_\text{thr} = 11.19~\text{GeV}^2$.

The nonvanishing cross section near threshold and the wide-range plateau have attracted great interest and driven a lot of theoretical studies. Some of the explanations put forward include a $B\bar{B}$ bound state or unobserved meson resonances~\cite{theory1}, Coulomb final-state interactions or quark electromagnetic interaction and scenarios that take into account the asymmetry between attractive and repulsive Coulomb factors~\cite{theory2}.
However, no obvious threshold effect was observed in the reactions $e^+e^-\rightarrow\Sigma\bar\Sigma$, $\Xi^{-}\bar\Xi^{+}$, and $\Xi^{0}\bar\Xi^{0}$~\cite{BESIII:2020uqk, BESIII:2021aer, BESIII:2020ktn}, indicating that the baryon antibaryon pair production in $e^+e^-$ annihilation near the threshold region is not fully understood.

In this article, we report a measurement of the Born cross section and the effective form factor for the process $e^{+}e^{-}\rightarrow \Omega^{-} \bar{\Omega}^{+}$ based on data samples collected at eight c.m.\ energies $\sqrt{s} = 3.4900$, 3.5080, 3.5097, 3.5104, 3.5146, 3.5815, 3.6500, and 3.6702 GeV with the BESIII detector~\cite{besiii} at the BEPCII~\cite{BESIII_YU}.

\section{BESIII Detector and Data Samples}
The BESIII detector~\cite{besiii} records symmetric $e^+e^-$ collisions provided by the BEPCII storage ring~\cite{BESIII_YU} in the center-of-mass energy range from
2.0 to 4.95~GeV~\cite{Ablikim:2019hff}, with a peak luminosity of $1 \times 10^{33} {\rm cm}^{-2}{\rm s}^{-1}$ achieved at $\sqrt{s} = 3.77\;\text{GeV}$.
The cylindrical core of the BESIII detector covers 93\% of the full solid angle and consists of a helium-based multilayer drift chamber~(MDC), a plastic scintillator time-of-flight(TOF)
system~, and a CsI(Tl) electromagnetic calorimeter~(EMC),
which are all enclosed in a superconducting solenoidal magnet
providing a 1.0~T magnetic field. The solenoid is supported by an octagonal flux-return yoke with resistive plate counter muon identification modules interleaved with steel~\cite{det:You}.
The charged-particle momentum resolution at 1 GeV/$c$ is $0.5\%$, and the specific energy loss (${\rm d}E/{\rm d}x$) resolution is $6\%$ for electrons from Bhabha scattering. The EMC measures photon energies with a resolution of $2.5\%$ ($5\%$) at $1$~GeV in the barrel (end cap) region. The time resolution in the TOF barrel region is 68~ps, while that in the end cap region is 110~ps.  The end cap TOF system was upgraded in 2015 using multigap resistive plate chamber technology, providing a time resolution of 60~ps~\cite{etof}.

To determine the selection efficiency,
200,000 $e^+e^-\rightarrow\Omega^{-}\bar{\Omega}^{+}$ signal MC events are generated evenly distributed in the phase space (PHSP) for each of the eight energy points using
the generator \textsc{conexc}~\cite{conexc}, which takes into account the beam-energy spread and corrections from initial-states radiation (ISR).
The $\Omega^{-}$ ($\bar{\Omega}^{+}$) hyperon is modeled with \textsc{evtgen}~\cite{evtgen} to decay inclusively according to the branching fractions taken from the Particle Data Group (PDG)~\cite{PDG2020}. The response of the BESIII detector is modeled
using a framework based on \textsc{geant}{\footnotesize 4}~\cite{geant4}. Large simulated samples of generic $e^+e^- \to \text{hadrons}$ events (inclusive MC)
are used to study possible background reactions with a generic event type analysis tool, TopoAna~\cite{TopoAna}.

\section{Event Selection}
As the selection of $e^{+} e^{-} \rightarrow \Omega^{-} \bar{\Omega}^{+}$ events with a full reconstruction of all six final-state particles
suffers from a low reconstruction efficiency, a single hyperon tag technique is applied and the events are reconstructed in one of three categories: (1) Only the $\Omega^{-}$ is reconstructed through the decay chain of $\Omega^{-}\to\Lambda K^{-}$ and $\Lambda\to p\pi^{-}$. The non-reconstructed $\bar{\Omega}^{+}$ is identified in the recoil against $\Omega^{-}$. (2) Only the $\bar{\Omega}^{+}$ is reconstructed and the $\Omega^{-}$ is missed. (3) Both $\Omega^{-}$ and $\bar{\Omega}^{+}$ are fully reconstructed. The three categories are mutually exclusive and there is no double counting for the reconstructed events.

Charged tracks are required to be reconstructed within the angular coverage
of the MDC: $|\!\cos\theta|<0.93$, where $\theta$ is the polar angle
with respect to the direction of the positron beam. The d$E$/d$x$ information obtained from MDC combined with
the flight time in the TOF is used to calculate the probability for the track being a pion, kaon or proton, respectively.
Each track is assigned to the particle type with the highest probability. Events with at least one negatively charged pion, one negatively charged kaon and one proton
are kept for further analysis.

The intermediate $\Lambda$ candidate is reconstructed by a secondary vertex fit~\cite{XUM} that is applied to all $p\pi^{-}$ combinations. 
The $p\pi^{-}$
invariant mass ($M_{p\pi^{-}}$) of the selected candidate is required to be within 4 MeV/$c^{2}$ of the nominal $\Lambda$ mass from the PDG~\cite{PDG2020}, determined by 3$\sigma$ of the $\Lambda$ peak from MC studies.
To further suppress background from non-$\Lambda$ events, the $\Lambda$ decay
length is required to be greater than zero. Here, the $\Lambda$ decay length is defined by the position vector from the primary to the secondary vertex, taking into account the crossing angle of the position vector with the momentum of $\Lambda$. After the above mentioned selection criteria it is possible that multiple $\Lambda$ candidates are reconstructed in one event; the best candidate selection is performed simultaneously to the $\Omega$ reconstruction as described in the following.
 \begin{figure}[!hbpt]
 \begin{center}
 \includegraphics[width=0.45\textwidth]{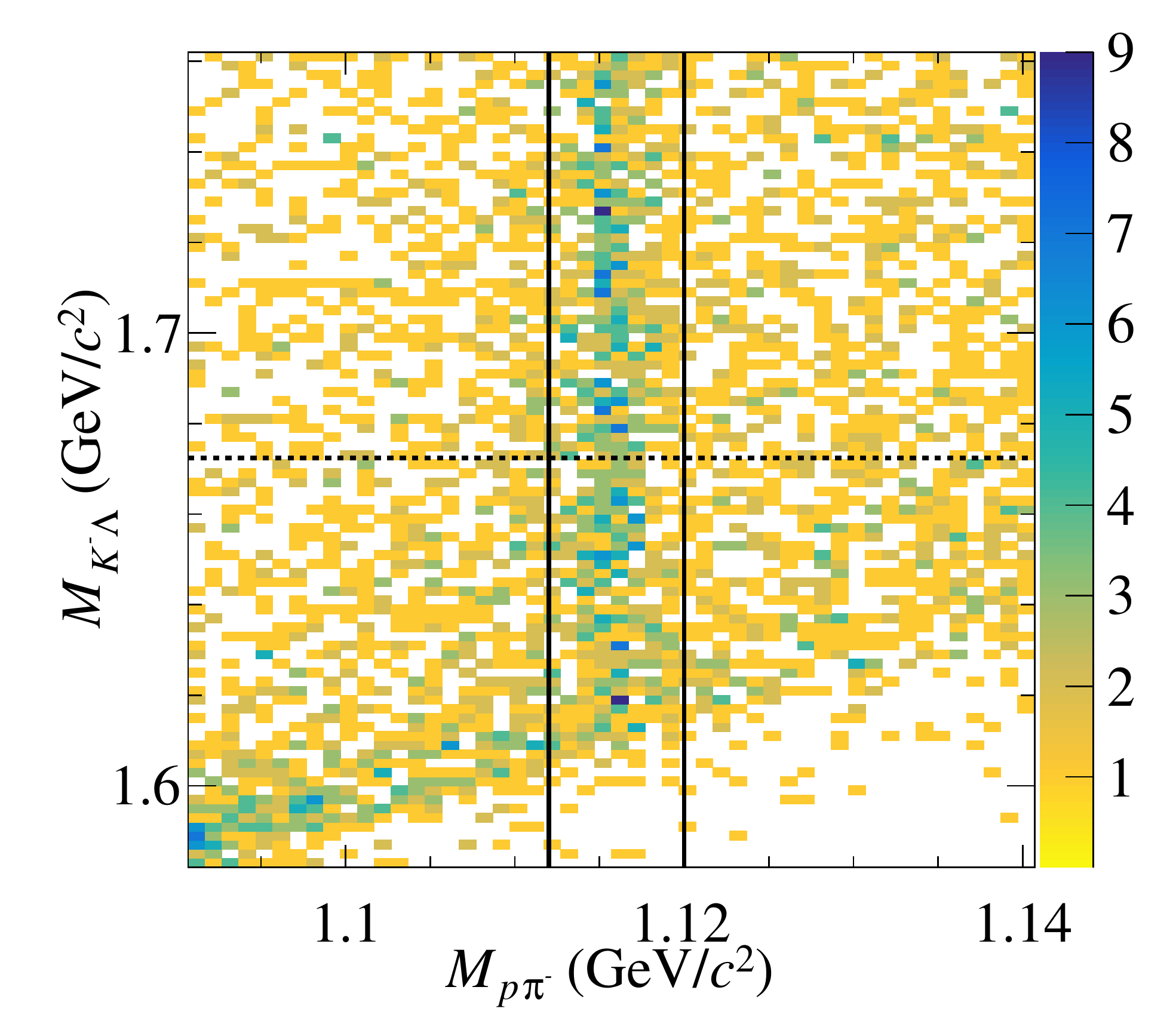}
 \end{center}
	 \caption{Two-dimensional distribution of $M_{K^{-}\Lambda}$ versus $M_{p\pi^-}$ from the sum of the eight energy points. The black solid lines indicate the $\Lambda$ window requirement, and the black dotted line indicates the $\Omega$ mass.}
 \label{mlam01}
 \end{figure}

The $\Omega^{-}$ candidate is reconstructed by combining a $\Lambda$ candidate with a  $K^{-}$. Similar to the $\Lambda$ reconstruction, a secondary vertex fit is applied. In cases with more than one $\Lambda$ or $\Omega^{-}$ candidate, the variable $\delta =(M_{p\pi^{-}}-m_\Lambda)^2 + (M_{K^{-}\Lambda}-m_{\Omega^-})^2$ is minimized to choose the best combination, where $M_{K^{-}\Lambda/p\pi^{-}}$ is the
invariant mass of the $K^{-}\Lambda/p\pi^{-}$ pair, and $m_{\Lambda/\Omega^{-}}$ is the mass of the $\Lambda/\Omega^{-}$ hyperon taken from the PDG~\cite{PDG2020}.
To suppress background events, the decay length of $\Omega^{-}$ is also required to be larger than zero. Figure~\ref{mlam01} shows the distribution of $M_{K^{-}\Lambda}$ versus $M_{p\pi^-}$ for the combined eight energy points. A clear accumulation is observed around the mass of the $\Lambda$.

To select $\bar\Omega^{+}$ hyperon candidates, we use the
distribution of the mass recoiling against the selected $K^{-}\Lambda$
system,
\begin{equation}
  RM_{K^{-}\Lambda} = \sqrt{(\sqrt{s}-E_{K^{-}\Lambda})^{2} - |\vec{p}_{K^{-}\Lambda}|^{2}},
\end{equation}
where $E_{K^{-}\Lambda}$ and $\vec{p}_{K^{-}\Lambda}$ are the energy and momentum of the selected $K^{-}\Lambda$ candidate in the c.m.\ system.

\begin{figure*}[hbpt]
\begin{center}
\includegraphics[width=0.95\textwidth]{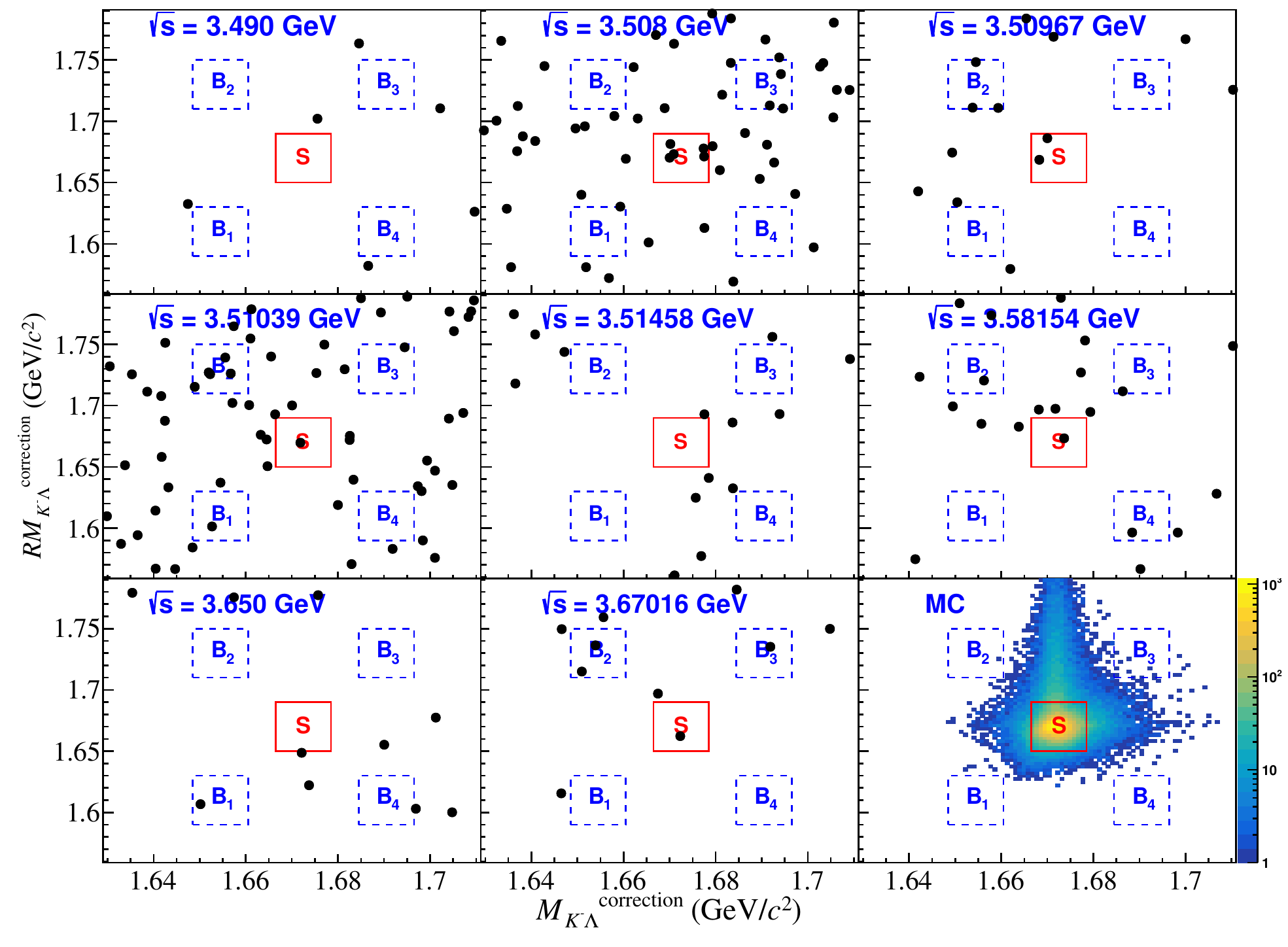}
\end{center}
\caption{Distribution of $RM^{\rm correction}_{K^{-}\Lambda}$ versus $M^{\text{correction}}_{K^{-}\Lambda}$ for eight energy points from data and the signal MC sample at 3.490 GeV (lower right plot). The red box denotes the $\Omega^{-}$ signal region, while the dashed blue boxes denote the sideband regions.}
\label{Fig:SP:DATA}
\end{figure*}

To improve the resolution, a correction is performed with $M_{K^-\Lambda}^{\text{correction}} = M_{ K^-\Lambda}-M_{p\pi^-}+m_{\Lambda}$.
Similarly, a correction is performed to the recoil side with $RM_{K^-\Lambda}^{\text{correction}} = RM_{K^-\Lambda}+M_{K^-\Lambda}-m_{\Omega^-}$.
Figure~\ref{Fig:SP:DATA} shows the distributions of the corrected recoil mass versus the corrected mass of $\Omega$ for each energy point, the central red box indicates the signal zone (zone S), while $ B_{1}, B_{2}, B_{3}, B_{4}$ marked by dashed blue boxes are for the considered sideband regions.

\section{Extraction of signal yields}
The signal yield for each energy point is extracted by counting the number of events in the signal and the sideband regions as shown in Fig.~\ref{Fig:SP:DATA}. The signal and background regions are defined as follows:
\begin{itemize}
 \item $S$: $M_{\Lambda K}\in [1.6665,1.6785]$, $RM_{\Lambda K}\in [1.65,1.69]$;
 \item $B_{1}$: $M_{\Lambda K}\in [1.6485,1.6605]$, $RM_{\Lambda K}\in [1.59,1.63]$;
 \item $B_{2}$: $ M_{\Lambda K}\in [1.6485,1.6605]$, $RM_{\Lambda K}\in [1.71,1.75]$;
 \item $B_{3}$: $ M_{\Lambda K}\in [1.6845,1.6965]$, $RM_{\Lambda K}\in [1.71,1.75]$;
 \item $B_{4}$: $ M_{\Lambda K}\in [1.6845,1.6965]$, $RM_{\Lambda K}\in [1.59,1.63]$.
\end{itemize}

Here, the window size for the tag and the recoil side are chosen as 3$\sigma$ of the mass spectrum from MC studies. The number of observed events in the signal region, $N_{\text{obs}}$, is determined. The number of background events in the sideband regions, $N_{\text{bkg}}$, is obtained by $N_{\text{bkg}} = \frac{1}{4}\sum_{i=1}^{4}N_{B_{i}}$, where $i$ runs over the four regions shown in Fig.~\ref{Fig:SP:DATA}. The number of signal events is obtained by
$N_{\text{S}} = N_{\text{obs}}-N_{\text{bkg}}$.
In Table~\ref{table:result}, the uncertainty of $N_{\text{S}}$ is calculated using the Feldman-Cousins method~\cite{FeldmanCousins}. Note that negative $N_{\text{S}}$ have been set to be 0 to avoid an unphysical number of signal events. Due to the low statistics, upper limits for these measurements are provided. The $N^{\text{UL}}$ is the upper limit at the 90\% confidence level, which is calculated with a profile likelihood method~\cite{TRolke, Lundberg:2009iu}. To investigate the potential background reactions, a study of the inclusive MC at 3.650 and 3.686 GeV is performed. Main background events are from non-$\Omega$ processes, such as $e^{+} e^{-} \rightarrow \Lambda \bar{\Lambda}\phi$ with $\phi \rightarrow K^+ K^-$. These background events can be estimated using the sideband strategy.

\section{Determination of Born cross section}
The Born cross section is calculated by
\begin{equation}
    \sigma^{B}(s)=\frac{N_{\text{S}}}{\mathcal{L}\cdot(1+\delta)\cdot \frac{1}{|1-\Pi|^{2}}\cdot\varepsilon} ,
\end{equation}
where $\mathcal{L}$ is the integrated luminosity, $(1+\delta)$ is the ISR correction factor, $\frac{1}{|1-\Pi|^{2}}$ is the vacuum polarization(VP) correction factor and $\varepsilon$ is the selection efficiency. The VP correction factor is obtained using the calculation described in Ref.~\cite{Kuraev:1985hb}. The ISR correction and efficiency $\varepsilon$ are determined using an iterative approach where a flat cross section line shape is adopted as an initial input and is iterated to obtain a stable result. In this analysis, the last two iterations have a difference of 0.03\% on the cross section.
The measured Born cross sections of the eight energy points are listed in Table~\ref{table:result}.

\begin{table*}[!hpt]
\begin{center}
\caption{\small
The numerical results for $e^+e^-\rightarrow\Omega^{-} \bar{\Omega}^{+}$. ${\cal L}$ is the integrated luminosity~\cite{chic1_data, 3650_lum}, $\frac{1}{|1 - \prod|^{2}}$ is the VP correction factor, $1 + \delta$ is the ISR correction factor,  $\epsilon$ is the selection efficiency, $N_{\text{obs}}$ denotes the number of observed events in the signal region, $N_{\text{bkg}}$ denotes the number of background events estimated with the sideband regions, $N_{\text{S}}$ ($N^{\text{UL}}_{\text{S}}$) is the number (upper limit) of signal events, $\sigma^{B}$ represents the Born cross section, and $G_{\text{eff}}(s)$ is the effective form factor. For the cross section and form factor, the first uncertainty is statistical and the second is systematic.}
\begin{tabular}{ccccccccc} \hline\hline
$\sqrt{s}$ (GeV) &  ${\cal L}$ (pb$^{-1})$ & $\frac{1 + \delta}{|1 - \prod|^2}$ & $\varepsilon$  &   $N_\text{obs}$  &$N_{\text{bkg}}$   &$N_{\text{S}}$($N^{\text{UL}}_{\text{S}}$)       & $\sigma^{B}$ (fb) & $|G_{\text{eff}}(s)|(\times 10^{-3})$ \\ \hline
3.4900  &  12.11  & 0.88  &  0.071  & 0 & 0 &  $0.0^{+1.3}_{-0.0}$ $(<2.0)$  &  $0^{+1780}_{-0}\pm 0$ $(<2738)$  &  $0^{+24}_{-0}\pm 0$ ($<$ 30) \\
3.5080  &  181.79 &0.89  &  0.075  & 5& 0.5 & $4.5^{+2.8}_{-2.2}$ $(< 9.6)$  &  $371^{+232}_{-185}\pm 16$ $(<797)$  &  $11^{+3}_{-3}\pm1$ $(< 16)$ \\
3.5097  &  39.29  &0.89  &  0.078  & 2&0.75 & $1.3^{+2.3}_{-0.9}$ $(< 5.3)$  &  $458^{+825}_{-339}\pm 20$ $(<1947)$  &  $12^{+8}_{-6} \pm 1$ $(< 25)$ \\
3.5104  & 183.64  &  0.89  &  0.077  & 1&1.5 &  $0.0^{+1.3}_{-0.0}$ $(<2.6)$  & $0^{+105}_{-0}\pm 0$ $(<231)$  &  $0^{+6}_{-0}\pm 0$ $(<9)$ \\
3.5146  & 40.92  & 0.89  &  0.080  & 0&0 & $0.0^{+1.3}_{-0.0}$ $(<2.0)$  &  $0^{+443}_{-0}\pm 0$ $(<682)$  &  $0^{+12}_{-0}\pm 0$ ($<$ 14) \\
3.5815  & 85.28 &  0.92  &  0.100  & 1&0.75 & $0.3^{+1.8}_{-0.3}$ $(<3.6)$  &  $32^{+225}_{-32}\pm 1$ $(<469)$  &  $3^{+6}_{-3}\pm1$ $(<11)$ \\
3.6500  &  44.49 & 0.92  &  0.120  & 0&0.25 &  $0.0^{+1.0}_{-0.0}$ $(<2.0)$  &  $0^{+214}_{-0}\pm 0$ $(<408)$  &  $0^{+7}_{-0}\pm 0$ $(<10)$ \\
3.6702  & 83.61 & 0.90  &  0.120  & 1&0.75 & $0.3^{+1.8}_{-0.3}$ $(<3.6)$  &  $29^{+200}_{-29}\pm 1$ $(<417)$  &  $3^{+5}_{-3}\pm 1$ $(<10)$ \\

\hline\hline
\end{tabular}
\label{table:result}
\end{center}
\end{table*}

\section{Determination of effective form factor}

For the spin-$\frac{3}{2}$ $\Omega^{-}$ hyperon, the effective form factor $|G_{\text{eff}}(s)|$ can be defined by a combination of the four form factors: $|G_{E0}|$, $|G_{M1}|$, $|G_{E2}|$ and $|G_{M3}|$~\cite{koerner,ramalho}.

\begin{equation}
\begin{aligned}\label{FF4000}
          |G_{\text{eff}}(s)| = \sqrt{\frac{2\times\frac{s}{4m^{2}}|G_{M}^{*}(s)|^{2}+ |G_{E}^{*}(s)|^{2}}{2\times\frac{s}{4m^{2}}+1}}   .
\end{aligned}
\end{equation}

Here, $m$ is the $\Omega^{-}$ mass, and $|G_{E}^{*}(s)|$ and $|G_{M}^{*}(s)|$ are defined as follows~\cite{ramalho}:
\begin{equation}
  \begin{aligned}
    |G_{E}^{*}(s)|^{2} &= 2|G_{E0}|^{2} + \frac{8}{9}\left(\frac{s}{4m^{2}}\right)^2|G_{E2}|^{2}, \\
    |G_{M}^{*}(s)|^{2} &= \frac{10}{9}|G_{M1}|^{2} + \frac{32}{5}\left(\frac{s}{4m^{2}}\right)^{2}|G_{M3}|^{2}.
    \label{eq:offs}
  \end{aligned}
\end{equation}

The total cross section of $e^{+}e^{-}\to\Omega^{-}\bar{\Omega}^{+}$ is related to the form factors as
\begin{equation}
  \sigma^{B}(s) = \frac{4\pi\alpha^{2}C\beta}{3s}\bigg[|G_{M}^{*}(s)|^{2} + \frac{2m^{2}}{s}|G_{E}^{*}(s)|^{2} \bigg],
  \label{eq:otxs}
\end{equation}
where $\alpha$ is the fine structure constant, $\beta = \sqrt{1-\frac{4m^{2}}{s}}$ is the velocity and $C = y/(1-e^{-y})$ is the Coulomb factor parametrizing the electromagnetic interaction between the outgoing baryon and antibaryon with $y = \pi \cdot\alpha\cdot\sqrt{(1-\beta^{2})}/\beta$.

Using Eqs.~(\ref{FF4000}) and~(\ref{eq:otxs}), the effective form factor can be determined from the total cross section by
\begin{equation}
\begin{aligned}
 |G_{\text{eff}}(s)| = \sqrt{\frac{\sigma^{B}(s)}{(1+\frac{2m^{2}}{s})\cdot(\frac{4\pi\alpha^{2}C\beta}{3s})}}.
\end{aligned}
\end{equation}

For the eight energy points in this analysis, the corresponding effective form factors are listed in Table~\ref{table:result}. The upper limits on the cross sections and form factors are determined using the profile likelihood method incorporating the systematic uncertainties, where the systematic uncertainties are also included in the profile likelihood method.

\section{Fit to Born cross section}
\begin{figure}[!hbpt]
 \begin{center}
 \includegraphics[width=0.48\textwidth]{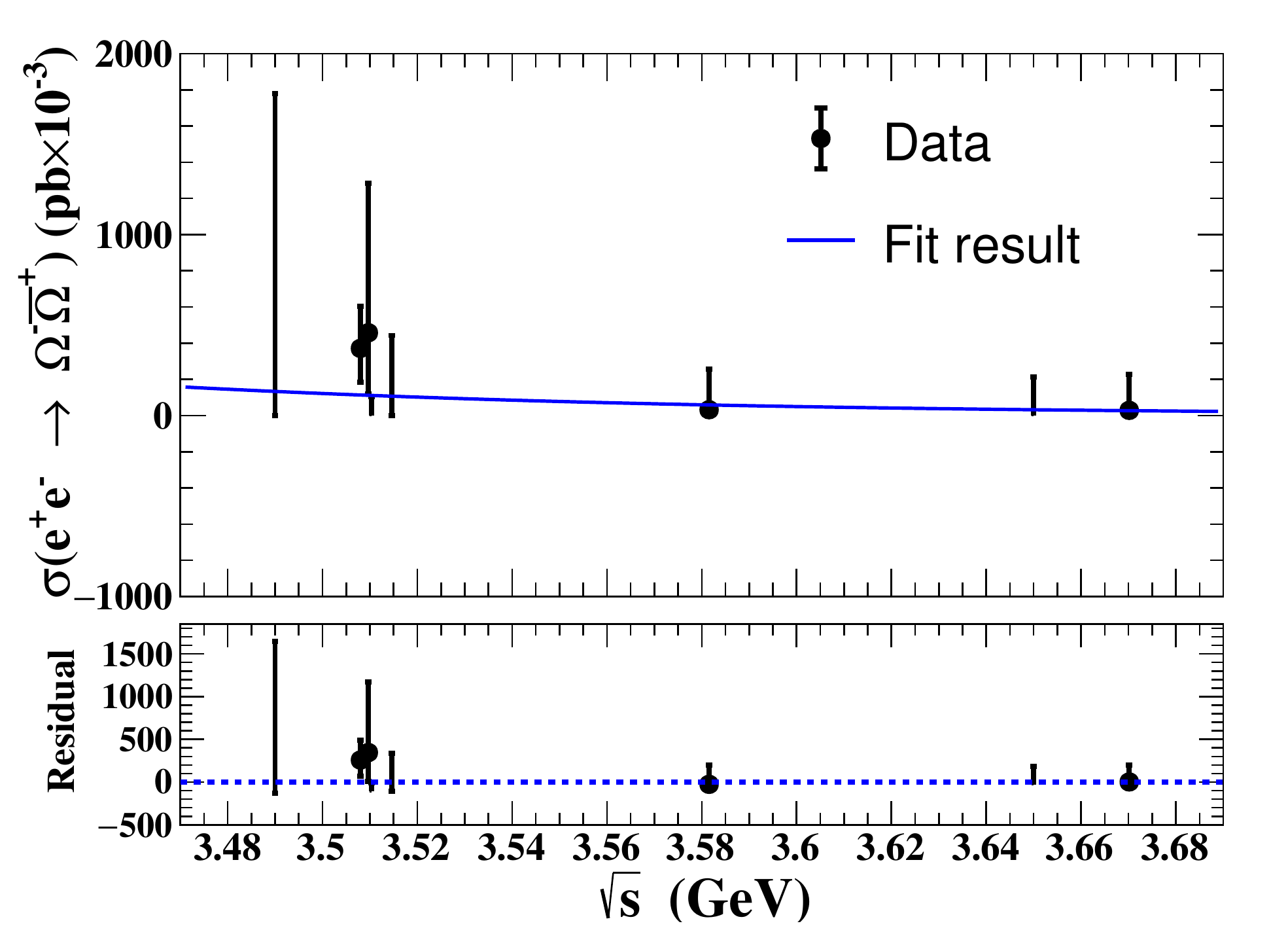}
 \end{center}
 \caption{Fit to the measured Born cross section with a pQCD-driven energy power function. The dots with the error bars are the measured Born cross section at c.m.\ energies between 3.49 and 3.67~GeV. The blue line denotes the fit results. The bottom plot shows the residual distribution.}
 \label{Fig:fit:threshold}
 \end{figure}

A least-$\chi^{2}$ fit to the Born cross section of  $e^{+} e^{-} \rightarrow \Omega^{-} \bar{\Omega}^{+} $ is performed with a perturbative QCD (pQCD) driven energy power law function,
\begin{equation}
\begin{aligned}
            \sigma^{B}(\sqrt{s}) = \frac{c_{0}\cdot\beta\cdot C}{(\sqrt{s}-c_{1})^{10}} ,
\end{aligned}
\end{equation}
where $c_0$ and $c_1$ are free parameters. This model has been applied in studies of
$e^+e^-\rightarrow \Lambda\bar\Lambda$~\cite{BESIII:2017hyw},  $\Xi^{0}\bar\Xi^{0}$~\cite{BESIII:2021aer}, and $\Xi^{-}\bar\Xi^{+}$~\cite{BESIII:2020ktn} production.
The fit returns $c_0 = (74.8^{+3103.1}_{-73.0})$~pb $\cdot$ GeV$^{-10}$ and $c_1 = (2.65 \pm 0.17)$~GeV, where both statistical and systematic contributions of the measured cross section have been included in the fit without considering the correlation between different energy points. Figure~\ref{Fig:fit:threshold} shows the fit result with quality $\chi^2/ndof =4.2/6.0$. The data can be well described by the pQCD driven energy power function, which indicates no obvious threshold enhancement.

\section{Systematic uncertainty}
Several sources of systematic uncertainties are considered on the Born cross section measurement. They include the
$\Omega$ reconstruction efficiency, the mass windows and decay length requirements of $\Lambda$ and $\Omega$,
sideband selection, angular distribution, the MC generator, and the fit models for the cross section.
Limited knowledge of the decay branching fractions of intermediate states and the luminosity measurement provide additional contributions. All the systematic uncertainties are discussed in detail below.
\begin{enumerate}
   \item The systematic uncertainty due to the $\Omega^{-}$ reconstruction efficiency, including the tracking and particle identification efficiency for the charged tracks and the efficiency for the $\Lambda$ reconstruction, is estimated by the control sample $\psi(3686) \rightarrow \Omega^{-}\bar{\Omega}^{+}$ with the same method as described in Refs.~\cite{BESIII:2012ghz, Ablikim:2016iym,BESIII:2019dve, BESIII:2021gca,BESIII:2021cvv, BESIII:2022mfx, BESIII:2022lsz}.
   \item The uncertainty associated with the mass windows of $\Lambda$ and $\Omega$ is estimated by changing the nominal window size from 3$\sigma$ to 4$\sigma$. The difference in results is taken as the uncertainty. The uncertainty from the decay lengths cut of $\Lambda$ or $\Omega^{-}$ is estimated by changing the nominal requirement of L$>$0 to an alternative requirement of L/$\sigma_L$ $>$ 2. Here $\sigma_L$ is the resolution of the decay length.
   \item The nominal sideband regions are defined as the windows of [6$\sigma$, 12$\sigma$] from the central value. By varying the position of the sideband window to [4.5$\sigma$, 10.5$\sigma$] or [7.5$\sigma$, 13.5$\sigma$], the maximum difference on the results is taken as the systematic uncertainty.
   \item In this analysis, the selection efficiency for $e^+e^-\rightarrow\Omega^{-}\bar{\Omega}^{+}$ is determined based on a PHSP model, which may differ from the real angular distribution. Due to the limited statistics, it is difficult to perform a study of the angular distribution in detail. Alternatively, we utilize the angular distribution from theoretical prediction near threshold ~\cite{Andrzej} to reproduce a MC sample, and take the efficiency difference between the signal MC samples and the alternative MC as the systematic uncertainty due to the $\Omega^{-}$ angular distribution.
   \item The precision of the ISR calculation in the MC generator is better than 1\%. Here we conservatively assign 1\% as the uncertainty of the radiative correction. Using ConExc, the input line shape is iteratively modified until the final cross section becomes stable. The last two iterations have a difference of 0.03\% on the cross section, which is negligible compared to the uncertainty of the MC generator.
   \item The fit to the line shape of the Born cross section has influence on the ISR factor and the selection efficiency. The input line shape is changed by $\pm1\sigma$, where $\sigma$ is taken from the fit result. The resulting change in ISR factor and selection efficiency is taken as the systematic uncertainty due to the fit on the line shape.
   \item The uncertainties associated with the branching fractions of the intermediate states $\Omega^{-}$ and $\Lambda$ are taken from the PDG~\cite{PDG2020}.
   \item The luminosity at each energy point is measured using Bhabha scattering events with an uncertainty about 0.6\%~\cite{chic1_data, 3650_lum}, which is taken as the systematic uncertainty.
\end{enumerate}
Table~\ref{table:sys} summarizes the various systematic uncertainties on the cross section measurements. Assuming all sources to be independent, the total systematic uncertainty is obtained by summing over the individual contributions in quadrature.
\begin{table}[!htp]
\centering
{\caption{\small Systematic uncertainty on the measurement of the Born cross section (\%). The uncertainties are the same for the eight energy points.}\label{table:sys}}
\begin{tabular}{cc}  \hline \hline
Source                                 &Value (\%)\\ \hline
$\Omega$ reconstruction                &3.7\\
 $\Lambda$ and  $\Omega$ mass windows             &0.4 \\
$\Lambda$ and  $\Omega$ decay lengths             &1.5 \\
Sideband                  &0.3 \\
Angular distribution                                  &0.8\\
MC generator                         &1.0  \\
Fit on the line shape                        &0.2  \\
Intermediate states                                  &0.9 \\
Luminosity                             &0.6 \\
Total                              &4.3  \\ \hline\hline
\end{tabular}
\end{table}

\section{Summary}
In summary, using $e^+e^-$ collision data corresponding to a total luminosity of 670 pb$^{-1}$ collected with the BESIII detector at BEPCII, the upper limits on the Born cross section and effective form factor for the process $e^+e^-\rightarrow\Omega^{-}\bar{\Omega}^{+}$ are measured by means of a single hyperon tag method, at eight c.m.\ energies between 3.49 and 3.67 GeV. The corresponding results are listed in Table~\ref{table:result}. After a fit to the cross section of $e^+e^-\rightarrow\Omega^{-}\bar\Omega^{+}$ with a pQCD driven energy function, as shown in Fig.~\ref{Fig:fit:threshold}, no significant threshold effect is observed near the $\Omega^{-}\bar\Omega^{+}$ threshold. Note that the lowest energy point in this work is still about 150 MeV above the threshold.

\begin{figure}[!hbpt]
 \begin{center}
 \includegraphics[width=0.48\textwidth]{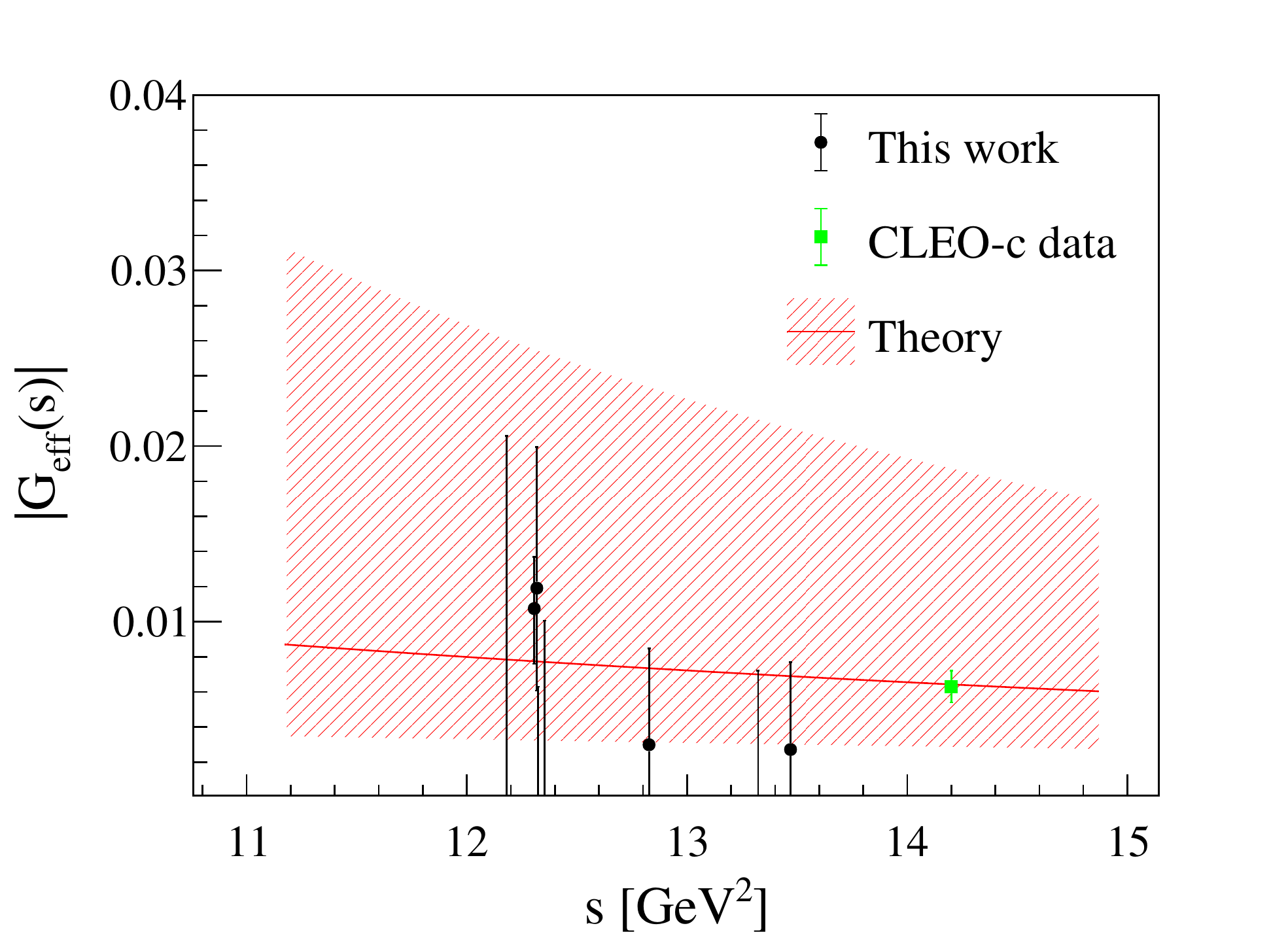}

 \end{center}
 \caption{
 Comparison of the effective form factor between this work, the CLEO-c data and the theoretical prediction~\cite{ramalho}. The black points with error bars are the results from this work. The green square is from a paper using the CLEO-c data~\cite{cleo} where $G_E = 0$ was assumed. The red band indicates the theoretical prediction with the red line in the middle showing the predicted central value~\cite{ramalho}, using CLEO-c data to fix the free parameters of the model. }
 \label{Fig:Geff}
 \end{figure}

The results of this analysis provide new experimental information to understand the production mechanism for hyperons with strangeness $S = -3$. The effective form factor determined in this paper (black data points) is consistent with the theoretical prediction (red band) using the covariant spectator quark model~\cite{ramalho} within the uncertainty of 1$\sigma$ as shown in Fig.~\ref{Fig:Geff}. Although only upper limits on the effective form factor are obtained due to the low statistics in this measurement, they will provide a constraint for the theoretical studies on the threshold effects of baryon pair production through $e^+e^-$ annihilation.

\section*{\boldmath ACKNOWLEDGMENTS}
\color{black}{}
The BESIII Collaboration thanks the staff of BEPCII and the IHEP computing center for their strong support. This work is supported in part by National Natural Science Foundation of China (NSFC) under Contracts 
   No. 11975278, No. 12075107, No. 12275320, No. 11635010, 
   No. 11735014, No. 11835012, N0.11905236, No. 11935015, No. 11935016, No. 11935018,
   No. 11961141012, No. 12022510, No. 12025502, No. 12035009, No. 12035013,
   No. 12047501, No. 12192260, No. 12192261, No. 12192262,
   No. 12192263, No. 12192264, No. 12192265, No. 12247101;
National Key Research and Development Program of China under Contracts No.  2020YFA0406400, No. 2020YFA0406300; the Chinese Academy of Sciences (CAS) Large-Scale Scientific Facility Program; Joint Large-Scale Scientific Facility Funds of the NSFC and CAS under Contracts No. U2032105, No. U2032109, No. U1832207; the CAS Center for Excellence in Particle Physics (CCEPP); 100 Talents Program of CAS; The Institute of Nuclear and Particle Physics (INPAC) and Shanghai Key Laboratory for Particle Physics and Cosmology; ERC under Contract No. 758462; European Union's Horizon 2020 research and innovation program under Marie Sklodowska-Curie grant agreement under Contract No. 894790; German Research Foundation DFG under Contracts No. 443159800, Collaborative Research Center CRC 1044, GRK 2149; Istituto Nazionale di Fisica Nucleare, Italy; Ministry of Development of Turkey under Contract No. DPT2006K-120470; National Science and Technology fund; National Science Research and Innovation Fund (NSRF) via the Program Management Unit for Human Resources and Institutional Development, Research and Innovation under Contract No. B16F640076; STFC (United Kingdom); Suranaree University of Technology (SUT), Thailand Science Research and Innovation (TSRI), and National Science Research and Innovation Fund (NSRF) under Contract No. 160355; The Royal Society, U. K. under Contracts No. DH140054, No. DH160214; The Swedish Research Council; U. S. Department of Energy under Contract No. DE-FG02-05ER41374.

\end{document}